\documentclass[
journal=jacsat,
layout=twocolumn,
manuscript=article
]{achemso}

\usepackage{chemformula} 
\usepackage[T1]{fontenc} 

\usepackage{xr}
\usepackage{mathrsfs,amsmath,amssymb}
\usepackage{graphicx}
\usepackage{multirow}
\usepackage{multicol}
\usepackage{booktabs}  

\usepackage{hyperref}
\usepackage{cleveref}
\crefname{figure}{\textcolor{black}{Figure}}{\textcolor{black}{Figure}}
\crefname{table}{\textcolor{black}{Table}}{\textcolor{black}{Table}}
\crefname{equation}{\textcolor{black}{Eq.}}{\textcolor{black}{Eq.}}
\crefname{section}{\textcolor{black}{Sec.}}{\textcolor{black}{Sec.}}

\author{Jing Chen}
\affiliation[Xiangtan University]
{Hunan Key Laboratory for Computation and Simulation in Science and Engineering, Key Laboratory of Intelligent Computing and Information Processing of Ministry of Education, School of Mathematics and Computational Science, Xiangtan University, Xiangtan, Hunan, China, 411105}

\author{Zhangpeng Sun}
\affiliation[Xiangtan University]
{Hunan Key Laboratory for Computation and Simulation in Science and Engineering, Key Laboratory of Intelligent Computing and Information Processing of Ministry of Education, School of Mathematics and Computational Science, Xiangtan University, Xiangtan, Hunan, China, 411105}

\author{Kai Jiang}
\affiliation[Xiangtan University]
{Hunan Key Laboratory for Computation and Simulation in Science and Engineering, Key Laboratory of Intelligent Computing and Information Processing of Ministry of Education, School of Mathematics and Computational Science, Xiangtan University, Xiangtan, Hunan, China, 411105}
\email{kaijiang@xtu.edu.cn}

\author{Jie Xu}
\email{xujie@lsec.cc.ac.cn}
\affiliation[Chinese Academy of Sciences]
{LSEC \& NCMIS, Institute of Computational Mathematics and Scientific/Engineering Computing (ICMSEC), Academy of Mathematics and Systems Science (AMSS), Chinese Academy of Sciences, Beijing, China.}

\title[An \textsf{achemso} demo]
{Investigation of double-gyroid grain boundaries beyond twinning}

\begin{document}
	
	%
	%
	%
	%
	%
	
	\begin{abstract}
		We study four double-gyroid (DG) grain boundaries (GBs) with different orientations numerically using the Landau--Brazovskii free energy, including the (422) twin boundary studied recently, a network switching GB, and two tilt GBs.
		Topological variations and geometric deformations are investigated. It is found that deviations in strut lengths and dihedral angles from the bulk DG substantially exceed changes in strut angles and nodal coplanarity. 
		We also examine the spectra along the contact plane of two grains and utilize them to evaluate the GB widths. 
		Of the four GBs we study, the network switching GB changes to the least extent topologically and geometrically, meanwhile has the lowest energy and the smallest GB width. 
	\end{abstract}

\section{Introduction}	
	The double gyroid (DG) has been observed in various materials, including block copolymers\,\cite{roy2011double, hajduk1994gyroid, meuler2009ordered, park2022double}, liquid crystal polymers\,\cite{molecular2020cao, otmakhova2021new}, lipid mesophases\,\cite{longley1983bicontinuous, aleandri2020physics}, surfactants\,\cite{tate2010how, sorenson2016unexpected, urade2007nanofabrication}, and biological assemblies\,\cite{michielsen2008gyroid, finnemore2009nanostructured, saranathan2010structure}. 
	The ideal DG skeleton has two interpenetrating networks with opposite chirality\,\cite{wang2020networks, prasad2018anatomy}, typically given by the higher-concentration region of a component. 
	Each network is featured by three coplanar struts of equal length from each node to three adjacent ones with the angle $120^{\circ}$. 
	Such planes rotate $\pm\arccos (1/3)$ ($\approx\pm 70.5^{\circ}$) along the clockwise($+$)/counterclockwise($-$) network, generating the shortest ten-node circuits\,\cite{feng2019topological} and the unit cell of the $Ia\bar{3}d$ space group. 
	The particular symmetries and skeletons lead to unique properties, including excellent mechanical\,\cite{dair1999mechanical, dair2000oriented, yang2019investigation, peng2021elastic, miralbes2022comparative, sharp2022corrosion}, optical\,\cite{dolan2014optical, hur2011three, saranathan2010structure, dolan2015optical}, mass transport properties\,\cite{ma2019mechanical, bobbert2017additively} and highly uniform porousness\,\cite{yanez2018gyroid, luo2020macroscopic}.
	These properties may be affected by defect structures\,\cite{khaderi2014stiffness} that are yet well-understood.
	
	Defects in network structures involve change of graph topology and geometry\,\cite{jinnai2001interfacial,li2014linking,feng2019topological,miyata2022dislocation, shan2023nature, dair1999mechanical, dair2000mechanical, feng2019seeing}, thus are more complicated. For DG, a well-structured $(422)$\,\footnote{Miller indices} twin boundary (TB) with mirror symmetry, possibly first noticed in Ref.\,\cite{vignolini2011a}, has been analyzed in several recent works.
	This specific structure is examined by minimal surface\,\cite{chen2018minimal, han2020crystal}, followed by experiments\,\cite{han2020crystal, feng2021visualizing}. 
	Specifically, the fusions of nodes on the TB plane and the resulting alterations of circuits are examined.
	To coordinate with the TB nodes, the network adjacent to these nodes undergo complex deformation, such as the changes of strut lengths and dihedral angles. 
	TBs are generally regarded to have lower energy because of their better symmetries, which is suggested by the results of spherical structures\,\cite{tschopp2015symmetric, li2019atomistic, bulatov2014grain, hahn2016symmetric}. 
	However, for DG and other network structures, it requires further studies on GBs of other orientations to validate.
	Furthermore, it is intriguing to explore whether some mechanisms on the formation of the $(422)$ TB also feature other GBs.
	
	In this work, we numerically examine four DG GBs of different relative orientations under the framework established in Ref.\,\cite{Xu2016Computing}, using the Landau--Brazovskii free energy that can describe many modulated phases including DG\,\cite{Brazovskii1975Phase,Fredrickson1987Fluctuation,Kats1993Weak,zhang2008efficient}.
	To briefly describe the framework, the GB structure is computed in a banded region between two parallel planes. The two grains with specific orientations are anchored outside, giving the boundary conditions on the two planes. 
	For the GBs we consider in this work, the two grains have common period parallel to the plane, so that a parallelogram unit cell in the plane is available. 
	The free energy is then minimized in a parallelepiped to obtain the optimal GB profile. 
	The numerical approach is established recently with high accuracy\,\cite{Cao2021Computing} that is adequately precise to resolve general anchoring orientations\,\cite{jiang2022tilt}. 
	
	We investigate the skeleton of the network using an isosurface of the GB profile. 
	The $(422)$ TB presented has the same topology with the experimental result in\,\cite{feng2021visualizing} and some differences in geometry. 
	More interestingly, we find that a network-switching (NetSw) GB turns out to have lower energy than TB, where one network is connected directly to the other with reversed chirality by struts. 
	We also examine a $[\bar{1}00]$ tilt GB (Tlt1) and a $[\bar{1}10]$ tilt GB (Tlt2), both with higher energy, where new nodes, curved struts are found and larger deformations are observed. 
	The deformation caused by GB is classified into four types: the differences of strut lengths, strut angles, dihedral angles from the bulk DG values, and the coplanarity of a node and its three adjacent ones.  
	The deformations in the two tilt GBs are indeed larger than those in the TB and the NetSw GB.
	The TB exhibits larger deformations on the contact plane, which may result in a higher energy than the NetSw GB. 
	
	Apart from the skeleton network, we also investigate the 2D spectra along the contact plane of the GBs that can be naturally extracted from the Fourier expansion. 
	By looking into the spectra with higher intensities, we can more effectively characterize the GB's constituents and more precisely define the GB widths, as previously suggested\,\cite{jiang2022tilt}. 
	Among the four GBs, the NetSw GB turns out to have the fewest number of spectra over certain intensity and the smallest width, which also suggests its lower energy. 
	
	\section{Methods}
	The Landau–Brazovskii (LB) free energy per volume of a scalar order parameter $\phi$, in a dimensionless form, is given by \,\cite{Brazovskii1975Phase,Fredrickson1987Fluctuation,Kats1993Weak}
	\begin{equation}\label{eq:energy}
		\begin{split}
			&E(\phi(\mathbf{r});\Omega)= \frac{1}{V(\Omega)} \int_\Omega \left\{\frac{1}{2}\left[\left(\nabla^{2}+1\right) \phi(\mathbf{r})\right]^{2} \right.\\
			& \quad\left. +\frac{\tau}{2}[\phi(\mathbf{r})]^{2}-\frac{\gamma}{3 !}[\phi(\mathbf{r})]^{3}+\frac{1}{4 !}[\phi(\mathbf{r})]^{4}\right\} \mathrm{d} \mathbf{r},
		\end{split}
	\end{equation}
	where $\mathbf{r}=(x,y,z)^T$, $V(\Omega)$ is the volume of the region $\Omega$, and the coefficients $\tau$, $\gamma$, depending on the material, may relate to various physical parameters \cite{leibler1980theory,ohta1986equilibrium}.
	Furthermore, the total mass 
	$\bar{\phi}=({1}/{V(\Omega)}) \int_\Omega \phi(\mathbf{r}) d \mathbf{r}$ is required to be conserved. 
	We choose $\tau=-0.4,\gamma=0.4$ in the DG region of the phase diagram.
	The bulk DG profile is obtained by minimizing the LB energy in a cubic unit cell using Fourier expansion, meanwhile optimizing the cell size\,\cite{zhang2008efficient,jiang2020efficient}. 
	
	To formulate the GB system, we pose the two grains with certain orientations and displacements, which can be done by rotating and shifting the bulk DG, in the half spaces $x<0$ and $x>0$, respectively. 
	Then, we choose an $L$ adequately containing the GB transition region and let GB relax in $-L\leq x \leq L$. 
	Anchoring boundary conditions are given by the function value and its normal derivative at $x=\pm L$, which are set equal to the (rotated and displaced) bulk values of two grains. 
	In the $y$-$z$ plane, we assume that two grains have common period and use 2D Fourier expansion according to the parallelogram unit cell. 
	Equivalently, from the spectral viewpoint, when the bulk DG profile is rotated, its spectra are rotated accordingly. 
	Hence, for the projections of rotated spectra of the two grains on the $y$-$z$ plane, the assumption of common period indicates that we could select two basis spectral vectors to express them with integer coefficients. 
	We shall emphasize that the boundary conditions and the basis spectral vectors depend on the posing of the two grains (See SI, \cref{sec:index}). 
	The discretization in the $x$-direction is implemented by generalized Jacobi polynomial that ensures sufficient accuracy\,\cite{Cao2021Computing}, and the energy minimization is solved by AA-BPG method\,\cite{jiang2020efficient}. 
	Numerical details can be found in previous works\,\cite{Cao2021Computing,jiang2022tilt}. 
	
	The GB energy per area is calculated by the difference between the free energy density of the GB system, $E_{gb}$, and that of the bulk, $E_0$, multiplied by the domain length in the $x$-direction, i.e., $\gamma_{g b}=2L \left(E_{g b}-E_0\right),$
	where $L$ is chosen as a multiple of the period of both (rotated) grains in the $x$-direction, and is sufficiently large to refrain finite-size effect.
	
	\section{Results and discussion}
	
	The orientations and displacements of two grains are specified by their contact plane at $x=0$ (supposing sharp interface, i.e. without relaxation), and further two coincident directions and one coincident point on the plane, as detailed in \cref{tab1}.
	Of the four GBs, it turns out that NetSw is the one with the lowest energy, while the two tilt GBs have higher energy. 
	We look into their structures below. 
	
	\begin{table}[t!]
		\centering
		\caption{The posing of two grains, where $a$ denotes the edge length of the DG unit cell.}
		\resizebox{\linewidth}{!}{
			\begin{tabular}{lcccccc}
				\toprule
				& & contact & \multicolumn{2}{c}{coincident} &  coincident & GB  \\
				& & plane & \multicolumn{2}{c}{directions} & point & energy\\
				\midrule
				\multirow{2}{*}{TB} & grain 1 & $(422)$ & $[1\bar{1}\bar{1}]$ & $[01\bar{1}]$ & $(0,0,a/2)$ & \multirow{2}{*}{$0.113$}\\
				& grain 2 & $(422)$ & $[\bar{1}11]$ &$[0\bar{1}1]$& $(0,0,a/2)$ \\
				\midrule
				\multirow{2}{*}{NetSw} & grain 1 & $(\bar{1}\bar{1}\bar{2})$ & $[\bar{1}\bar{1}1]$ & $[\bar{1}10]$ & $(0,0,0)$ & \multirow{2}{*}{$0.098$}\\
				& grain 2 & $(112)$ & $[\bar{1}\bar{1}1]$ & $[1\bar{1}0]$ & $(0,0,0)$\\
				\midrule
				\multirow{2}{*}{Tlt1} & grain 1 & $(0\bar{1}\bar{3})$ & $[0\bar{3}1]$ &$[\bar{1}00]$& $(0,0,0)$ & \multirow{2}{*}{$0.210$}\\
				& grain 2 & $(01\bar{3})$ & $[0\bar{3}\bar{1}]$ &$[\bar{1}00]$& $(0,0,0)$\\
				\midrule
				\multirow{2}{*}{Tlt2} & grain 1 & $(\bar{1}\bar{1}\bar{2})$ & $[\bar{1}\bar{1}1]$ &$[\bar{1}10]$& $(0,0,0)$ & \multirow{2}{*}{$0.273$}\\
				& grain 2 & $(11\bar{2})$ & $[\bar{1}\bar{1}\bar{1}]$ &$[\bar{1}10]$& $(0,0,0)$\\
				\bottomrule
		\end{tabular}}
		\label{tab1}
	\end{table}
	
	\subsection{Network Structure}\label{sec:real-space}
	To visualize the skeleton of a GB, we present the isosurface at $\phi=1.6$. 
	Nodes are identified at local maxima whose values are not less than $84\%$ of maximum bulk value. 
	The left-handed and right-handed networks are colored with blue and, respectively.
	
	\subsubsection{TB}
	\begin{figure}[tbhp]
		\centering
		\includegraphics[width=1\linewidth]{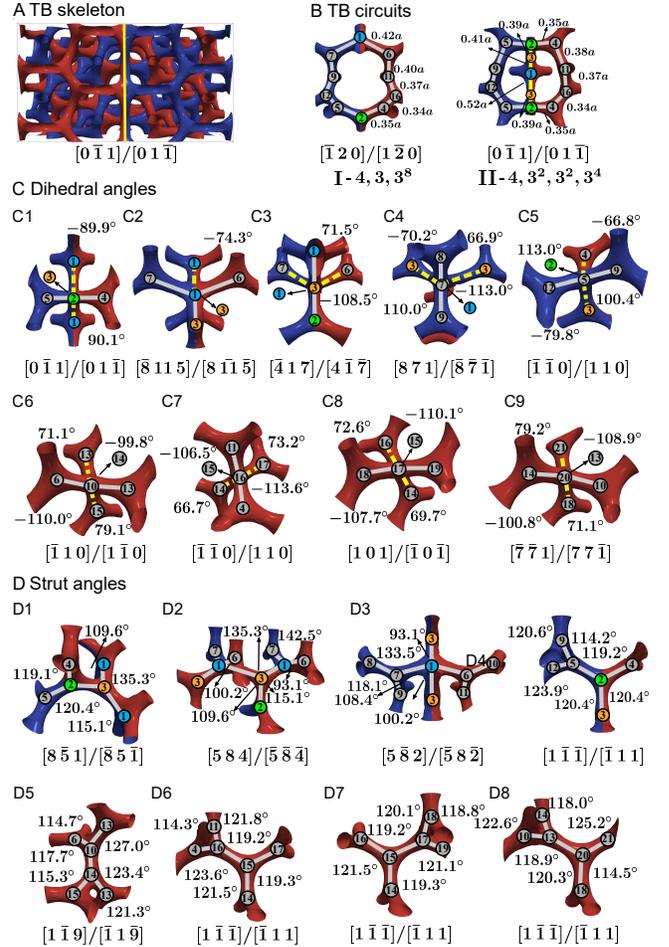}
		\caption{The TB skeletons. The viewing direction with respect to the left/right grain is labeled below each figure. 
			(A) The whole skeleton. 
			(B) Two circuits. Nodes with whole periods differences are labeled the same.
			(C) Dihedral angles, viewing along a strut direction. Solid lines are closer, dashed lines farther. 
			(D) Strut angles.}
		\label{twin.dihedral}
	\end{figure}
	The skeleton of TB is shown in \cref{twin.dihedral}, with two viewing directions relative to two grains displayed. 
	We affirm the mirror symmetry of TB after verifying that $|\phi(x,y,z)-\phi(-x,y,z)|<10^{-10}$. 
	We label the nodes with numbers, and use the same number to represent equivalent sites in different unit cells. 
	The new nodes generated on $x=0$ in TB are numbered with 1, 2, 3, where node 1 has four edges and node 2, 3 both have three edges.
	Notably, node 3 connects only to nodes on $x=0$ (two node 1 and one node 2, see D1), whereas node 1 and node 2 connect to two nodes outside $x=0$.  
	When TB nodes are not involved, the network topology of TB is identical to that of bulk DG. 
	Circuits involving TB nodes are divided into category $\rm \uppercase\expandafter{\romannumeral1}$ if nodes in the two grains are connected by TB nodes, and category $\rm \uppercase\expandafter{\romannumeral2}$ if circuits in two grains only share TB nodes (see \cref{twin.dihedral}\,B and SI,\,\cref{twin.loop}).
	We further specify the number of TB nodes and bulk nodes in the circuits. For instance, $\rm \uppercase\expandafter{\romannumeral1}\,\mbox{-}\,4,3,3^8$ comprises one node 1, one node 2, and eight bulk nodes, while circuit $\rm \uppercase\expandafter{\romannumeral2}\,\mbox{-}\, 4,3^2,3^2,3^4$ consists of a single node 1, two node 3, two node 2 and four bulk nodes.
	The network topology is consistent with that given in Ref.\,\cite{feng2021visualizing}.
	
	Substantial deformations occur when three TB nodes are involved. 
	Notably, struts attached to node 1 are elongated over $20\%$ from $0.35a$ (\cref{twin.dihedral}\,B), and vertical planes appear (\cref{twin.dihedral}\,C1). 
	The chirality of network is locally reversed as indicated by the dihedral angles (\cref{twin.dihedral},\,C2 and C4). 
	However, the value of dihedral angles are surprisingly close to $70.5^{\circ}$, which is different from what is reported previously \cite{feng2021visualizing}. 
	Large deviations of strut angles involving node 1 are also observed (\cref{twin.dihedral}\,D1-D3). 
	
	In bulk DG, a node and its three adjacent nodes are coplanar. 
	This seems to be well kept in TB when only one node on $x=0$ is involved, as the dihedral angles 8-(1,7)-9 and 9-(2,5)-12 almost equal $180^{\circ}$ (\cref{twin.dihedral},\,C4 and C5, the sum of two adjacent dihedral angles). 
	
	Looking into the geometry, we notice that the deviations of strut lengths and dihedral angles are larger than those of strut angles (see \cref{twin.dihedral},\,B, C6-C9 and D5-D8 and SI,\,\cref{twin.loop}). 
	Deformations are generally larger when nodes connected to $x=0$ are involved, for example node 4 and 6 (\cref{twin.dihedral},\,B, C6-C7, D5-D6), particularly in the elongation of struts (6-11 of the length $0.40a$) and deflection of coplanarity (see dihedral angles 4-(15,16)-11 and 6-(10,14)-13). 
	In contrast, for nodes not connected to $x=0$, coplanarity is almost maintained (for instance, see dihedral angles 13-(14,10)-15, 14-(15,16)-17, 18-(15,17)-19 and 18-(13,20)-21 in \cref{twin.dihedral}\,C6-C9), with a few exceptions (an example is 10-(13,20)-14 in \cref{twin.dihedral}\,C9). 
	Later, we shall analyze the statistics on changes of strut lengths, strut angles, dihedral angles and coplanarity. 
	
	\subsubsection{NetSw}\label{sec:hand}
	\begin{figure*}[h!]
		\centering
		\includegraphics[width=1\textwidth]{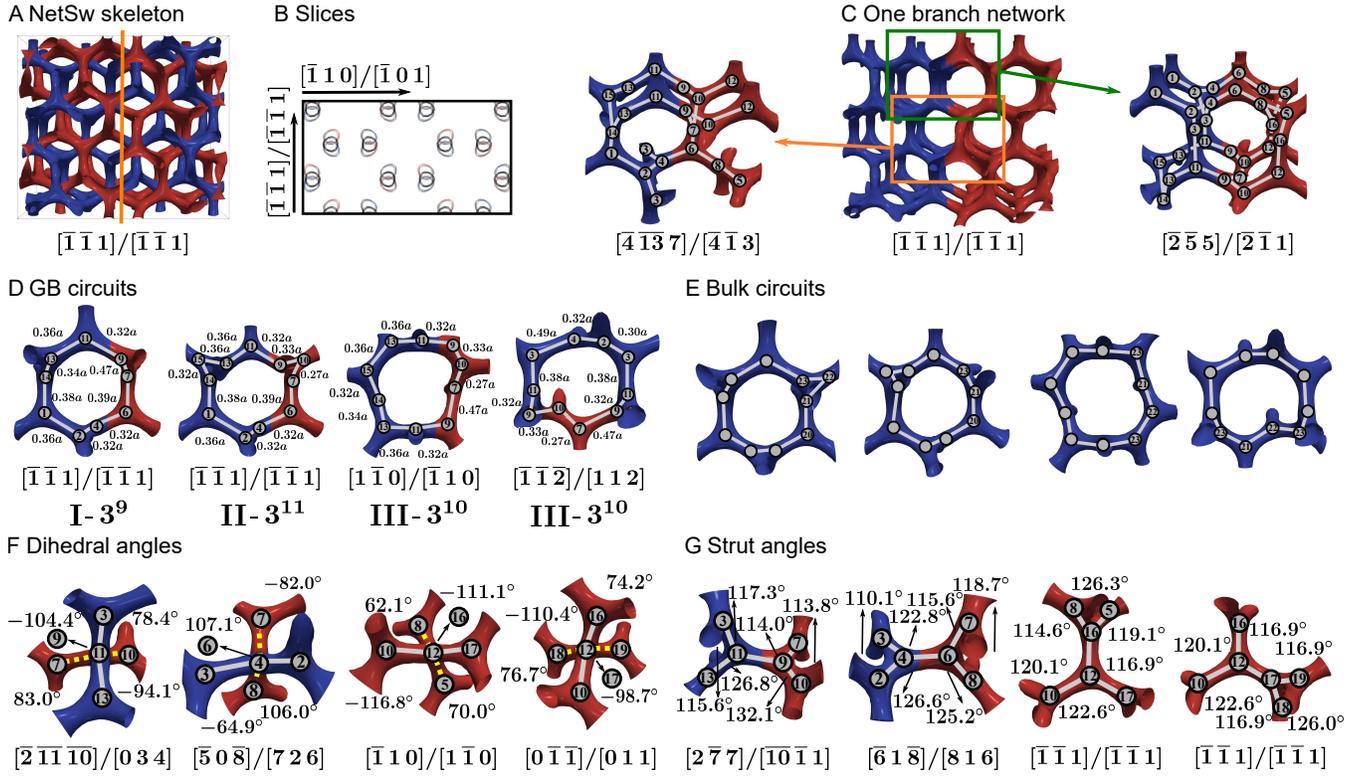}
		\caption{(A) The NetSw skeleton.
			(B) Slices of GB (black), grain 1 (blue) and 2 (red) on the plane $x=0$.
			(C) One branch of the network.
			(D) Circuits in GB and strut lengths, in comparison with (E) Bulk circuits of grain 1. 
			(F)(G) Dihedral angles and strut angles.
		}
		\label{hand}
	\end{figure*}
	
	The posing of two grains in NetSw suggests that it has a $180^{\circ}$ rotational symmetry around the $[\bar{1}\bar{1}1]$ axis of both grains (the $y$-axis in the computation). 
	Indeed, we confirm it by verifying that  $|\phi(x,y,z)-\phi(-x,y,-z)|<7\times10^{-5}$. 
	The skeleton of NetSw is shown in \cref{hand}\,A. 
	The networks with different chirality are connected through struts at the plane $x=0$ without generating new nodes, reflected by the slices of two grains before relaxation (\cref{hand}\,B, blue and red) and GB (black). 
	Due to the symmetry, we focus on one branch (left blue, right red, \cref{hand}\,C), where we shall remember that the different nodes may share the same number if they are whole periods different. 
	The circuits passing $x=0$ are formed by combining part of left- and right-handed ones, including nine, ten and eleven nodes (\cref{hand}\,D).
	These circuits have more blue nodes, each of which has a counterpart with more red nodes (SI,\,\cref{hand.ring}). 
	To comprehend these circuits, we compare them with the bulk circuits of grain 1 (\cref{hand}\,E). 
	It is observed that red nodes 6 and 9 are close to blue nodes 20 and 23 in the bulk, respectively. 
	But the connection between node 6 and 9 is altered due to reverse chirality, resulting in circuits with different lengths. 
	It also causes significant skeleton distortions. 
	While the struts crossing $x=0$ (4-6 and 9-11) deviate less than $10\%$ from $0.35a$, we observe huge elongations (3-4, 7-9) and shrinkages (7-10) in the vicinity of $x=0$. 
	The alterations in dihedral angles (\cref{hand}\,F) and strut angles (\cref{hand}\,G) are also larger near $x=0$, while strut angles generally deviate less than dihedral angles. 
	Meanwhile, by checking the sum of adjacent dihedral angles, we notice that the deflection of coplanarity appears to be much less than the change of dihedral and strut angles. 
	This observation is similar to TB, which we shall revisit later. 
	
	\subsubsection*{Tlt1}\label{sec:tilt}
	\begin{figure*}[h!]
		\centering
		\includegraphics[width=1\textwidth]{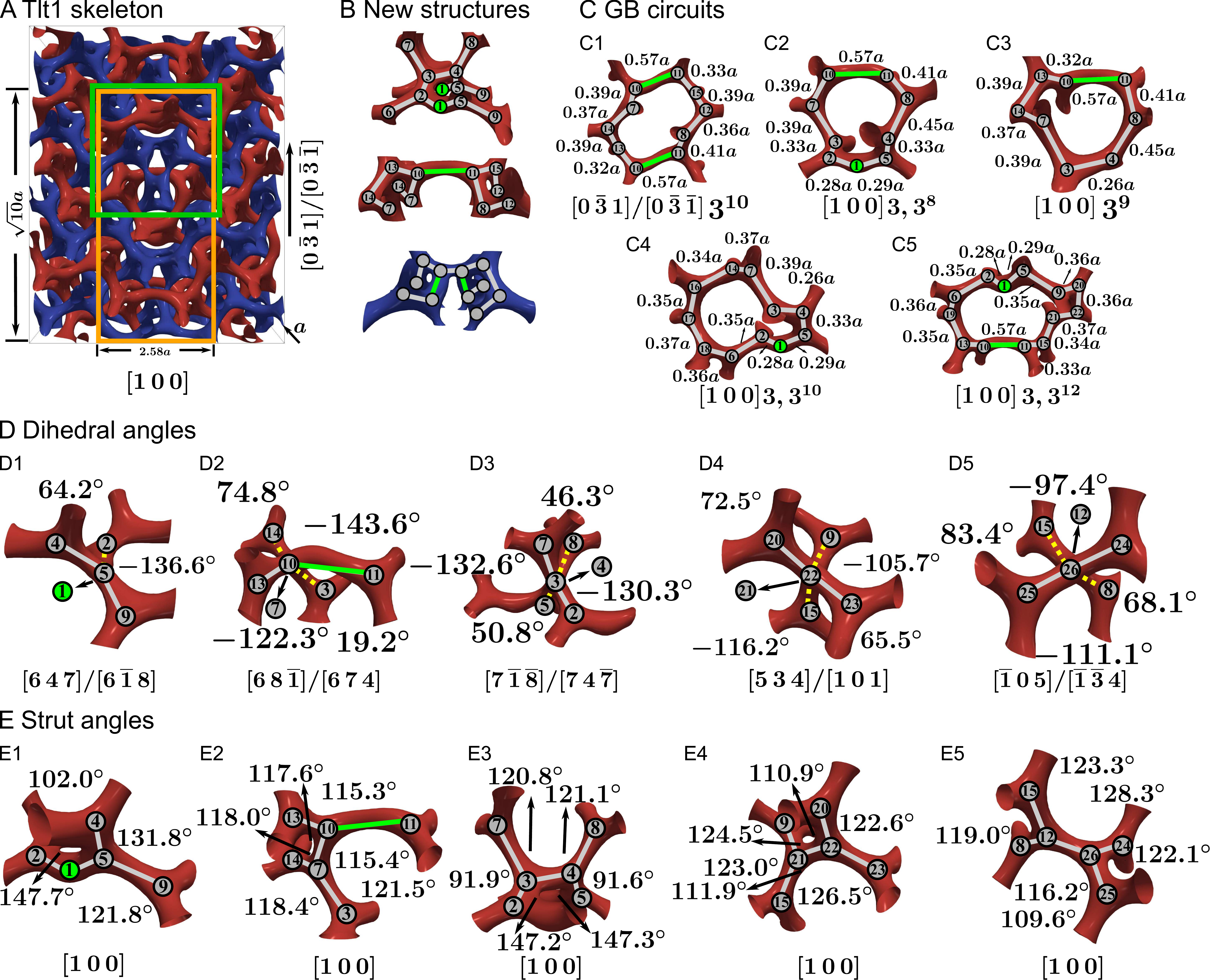}
		\caption{(A) Tlt1 skeleton. 
			The orange box encloses one period.
			(B) One new GB node and three curved struts. 
			(C)(D)(E) Circuits, dihedral angles and strut angles.
		}
		\label{tilt}
	\end{figure*}
	
	The skeleton of Tlt1 is presented in \cref{tilt}\,A. For both grains, the $[\bar{1}00]$ direction coincides with the $z$-axis, and the period length in the $y$-direction is $\sqrt{10}a$ (the height of the orange box in \cref{tilt}\,A). 
	We particularly focus on the green box where new node and curved edges appear, drawn in detail in \cref{tilt}\,B. 
	In Tlt1, networks of the same chirality for both grains are connected. 
	We focus on the red network and leave the blue to SI,\,\cref{tilt.ring.shape.2} since their features are similar. 
	
	We find one new node (number 1) and several curved edges between nodes (green ones in \cref{tilt}\,B), which resemble the so-called `network breaks' topological defects \cite{feng2019topological} or `point defects' \cite{miyata2022dislocation}.
	The other nodes can be regarded as displaced from bulk nodes, forming circuits crossing $x=0$ (\cref{tilt}\,C) by combining part of left and right grains at node 1, edge 3-4 and 10-11. 
	Strut lengths involving these nodes (1, 3, 4, 10, 11) connecting two grains deviate larger than others. 
	The dihedral angles and strut angles clearly vary more drastically than TB and NetSw (\cref{tilt},\,D and E), even far away from $x=0$ (\cref{tilt}\,D4,\,D5,\,E4,\,E5). 
	However, we notice that for many nodes not concerning node 1 and curved edges, coplanarity is almost maintained, such as 2-(3,4)-7, 5-(3,4)-8, 20-(21,22)-23 and 24-(12,26)-25. 
	
	\subsubsection{Tlt2}
	The Tlt2 has a skeleton more complicated, for which we present the topology in SI, Tlt2.
	Specifically, there are three new types of nodes, five curved struts, and twenty-four GB circuits that differ from the bulk circuits (see SI,\,\cref{tilt0.real}-\cref{tilt0.ring}).
	
	\subsubsection*{Deformation Statistics}\label{sec:shape}
	\begin{table}[h!]
		\centering
		\caption{Maximum variations for strut lengths, dihedral angles, strut angles, coplanarity.}
		\resizebox{\linewidth}{!}{
			\begin{tabular}{cccccc}
				\toprule
				\multirow{2}{*}{GBs} & \multirow{2}{*}{region} & strut length & strut angle  &  dihedral angle & coplanarity \\
				&  & $(0.35a)$ & $(120^{\circ})$  &  $(\pm70.5^{\circ})$ & $(180^{\circ})$\\
				\midrule
				\multirow{3}{*}{TB} & I & $\pm 0.20a$ & $\pm 27^{\circ}$ & $ \pm 43^{\circ} $ & $ \pm 45^{\circ}$ \\
				& A & $\pm 0.05a$ & $\pm 9^{\circ}$ & $ \pm 17^{\circ} $ & $ \pm 13^{\circ}$ \\
				&  B & $ \pm 0.05a$ & $ \pm 8^{\circ}$ &$ \pm 11^{\circ}$  & $ \pm 6^{\circ}$ \\ 
				\midrule
				\multirow{2}{*}{NetSw} & A & $\pm0.14a$ & $ \pm 12^{\circ}$ &  $ \pm 18^{\circ}$  & $ \pm 11^{\circ}$ \\
				&  B  & $\pm0.08a$ & $ \pm 10^{\circ}$ &$ \pm 13^{\circ}$ & $ \pm 9^{\circ}$\\
				\midrule
				\multirow{2}{*}{Tlt1} & A & $\pm 0.22a$ & $ \pm 28^{\circ}$& $ \pm 53^{\circ}$  & $ \pm 38^{\circ}$ \\
				& B & $\pm 0.05a$ & $ \pm 17^{\circ}$ & $ \pm 20^{\circ}$  & $ \pm 25^{\circ}$ \\
				\bottomrule
		\end{tabular}}
		\label{tab2}
	\end{table}
	
	\begin{figure*}[h!]
		\centering
		\includegraphics[width=1\textwidth]{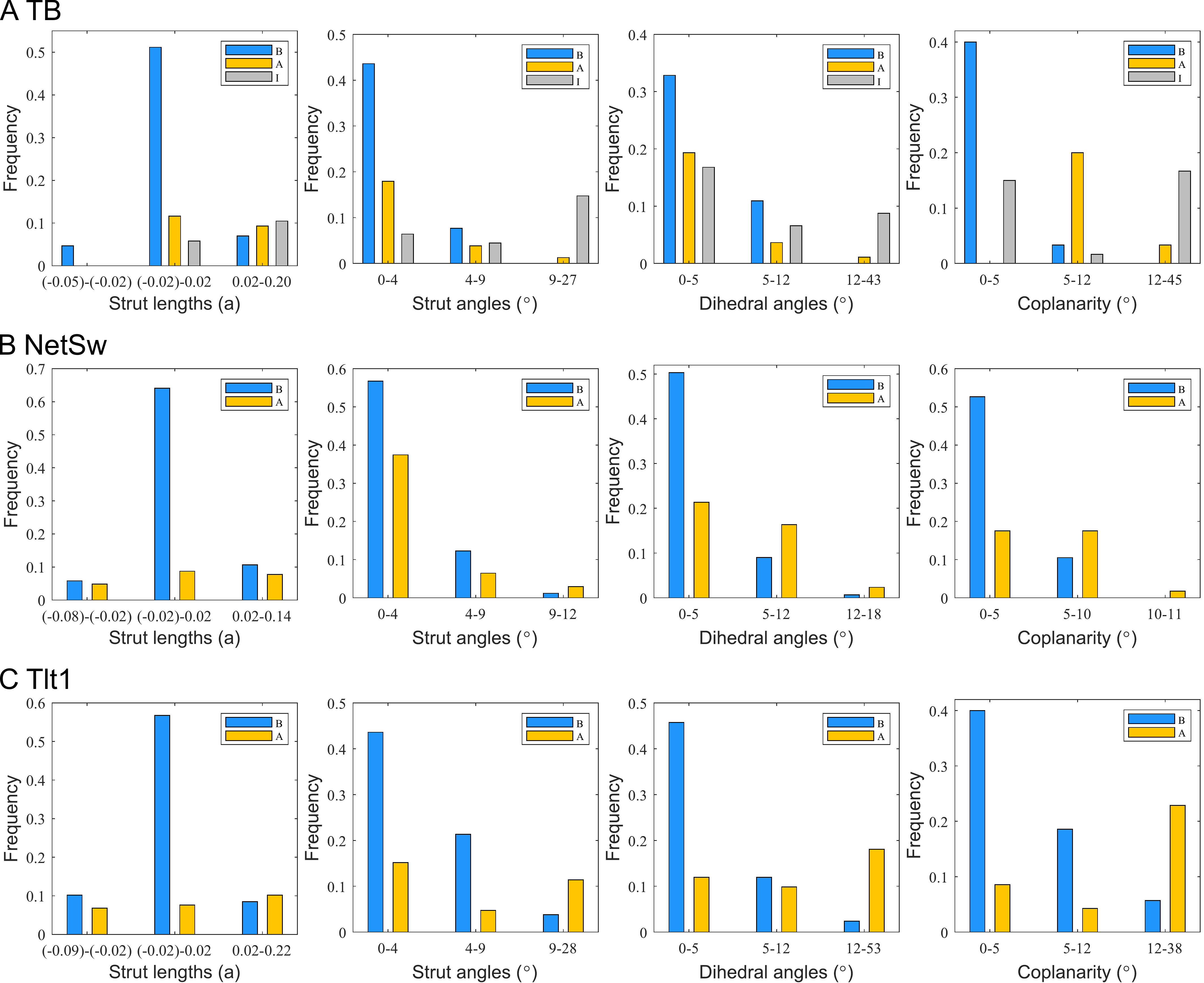}
		\caption{Statistics of deviations for strut lengths, strut angles, dihedral angles and coplanarity.}
		\label{data}
	\end{figure*}    
	
	We investigate the statistics of deviations for four geometric parameters (strut length, strut angle, dihedral angle and coplanarity) within $x\in[-1.58a, 1.58a]$ that covers the range of deformation for the four GBs examined in the current work. 
	Specifically, for the coplanarity of one node and three adjacent nodes, we examine the maximal deviation of dihedral angles formed by three planes, each determined by the node and its two connecting nodes.
	The nodes are categorized into three regions: region I comprises nodes on the plane $x=0$; region A includes nodes with a strut touching or crossing $x=0$; the remaining nodes reside in region B. 
	Accordingly, a strut is considered to be in region I if either of its nodes is, in region A if either node is in region A but neither is in region I, and in region B otherwise.
	The angles are categorized similarly. 
	
	The maximum deviations for TB, NetSw and Tlt1 are listed in \cref{tab2}, and histograms are presented in \cref{data}. 
	The histogram intervals are chosen approximately at $3\%, 7\%$ for strut angles (w.r.t. $120^{\circ}$) and coplanarity (w.r.t. $180^{\circ}$), while $7\%$ and $17\%$ for dihedral angles (w.r.t. $70.5^{\circ}$). 
	For strut lengths, the middle interval is selected to span approximately $6\%$ above and below the bulk value of $0.35a$.
	The range and statistics for Tlt2 are provided in SI,\,\cref{data_tilt0} and \cref{tab_tilt0}. 
	
	The statistics clearly indicate that Tlt1 (as well as Tlt2, see SI) deforms greater than TB and NetSw in all four aspects, potentially contributing to its higher GB energy. 
	Focusing on TB and NetSw, it is evident that huge distortions in TB mostly occur in region I, while the distortions in region A and B might be slightly less than NetSw. 
	It suggests that region I contributes significantly on higher GB energy of TB than NetSw. 
	Furthermore, we are also interested in the comparison of four geometric parameters in Region A and B of TB and NetSw.
	It is noticed that strut lengths and dihedral angles are more prone to altering (many exceeding $7\%$) than strut angles and coplanarity (most within $7\%$).
	In particular, dihedral angles appear to change the most but coplanarity is maintained better, which is consistent with our intuitions obtained from \cref{twin.dihedral}-\cref{tilt}. 
	This observation implies that coplanarity may play an essential role in GB energy.
	
	\subsection{Spectral Configurations}\label{sec:modes}
	\begin{figure*}[h!]
		\centering
		\includegraphics[width=1\textwidth]{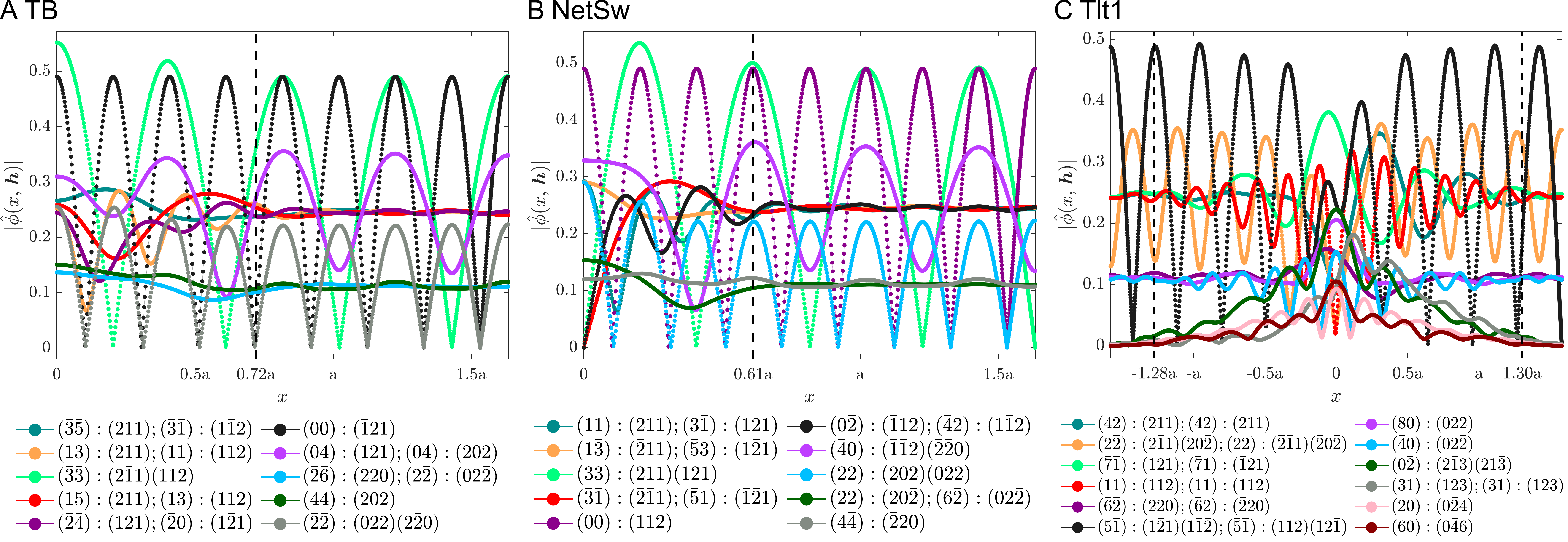}
		\centering
		\caption{Intensities of spectra against $x$, for those maximum over $0.09$. (A) TB, (B) NetSw and (C) Tlt1. 
			The black dotted line represents the evaluated boundary of GBs using \cref{width} with $\alpha=0.03$.
			The correspondence of bulk indices and 2D projection indices are provided in the legends, where we only list bulk indices with higher intensities in the bulk DG profile. 
		}
		\label{mode}
	\end{figure*}
	
	Since the GB profile $\phi$ is periodic in the $y$-$z$ plane, we could examine its spectral intensities (i.e. moduli of Fourier coefficients, denoted by $|\hat{\phi}(x,\boldsymbol{k})|$ for certain index $\boldsymbol{k}$) as functions of $x$.
	Each spectrum is the projection of multiple bulk GB spectra onto the $y$-$z$ plane. 
	Thus, it is natural to focus on the projection of major bulk indices, i.e. the two classes $\{211\}$ and $\{220\}$. 
	In particular, we investigate those spectra with maximum intensity within $x\in[-L,L]$ exceeding $0.09$, which is approximately $36\%\,(80\%)$ of the intensity of $\{211\}\,(\{220\})$ in the bulk DG. 
	For the GBs studied in this work, other spectra have the intensities sufficiently smaller. 
	The intensities of the chosen spectra are plotted in \cref{mode}, where for TB and NetSw, only the positive $x$ is presented due to mirror symmetry with respect to $x=0$ (but it is not the case for Tlt1). 
	Furthermore, intensities of different spectra may coincide, so that one curve may represent multiple spectra. 
	The spectral intensities of Tlt2 is provided in SI,\,\cref{intensity0}. 
	The correspondence between the 2D indices in GB and the projected 3D indices in bulk DG is also given (dependent on the orientations of grains) in \cref{mode} (see SI,\,\cref{tab: mode indices} for details).
	
	Let us first look into the spectral indices. 
	The number of chosen spectra (with maximum intensity exceeding $0.09$) is 16, 15, 19 and 17 for TB, NetSw, Tlt1 and Tlt2, respectively, of which NetSw is the smallest.  
	Specifically, for TB and NetSw, only projections of major bulk spectra are present. 
	Interestingly, in NetSw, these projections occupy fewer sites in the $y$-$z$ plane, which may lead to lower GB energy.
	Conversely, several additional spectra emerge for Tlt1 and Tlt2, which may be closely related to higher GB energy. 
	
	To have a closer inspection of the spectral intensities, we compare with the bulk spectral intensities in the $y$-$z$ plane, defined as follows. 
	Consider two grains (with prescribed orientations and displacements) occupying the two half planes $x<0$ and $x>0$. 
	Since they are also periodic in the $y$-$z$ plane, we can obtain the spectral intensities, denoted as $|\hat{\phi}_{\mathrm{bulk}}(x,\boldsymbol{k})|$. 
	The GB spectral intensities $|\hat{\phi}(x,\boldsymbol{k})|$ resemble $|\hat{\phi}_{\mathrm{bulk}}(x,\boldsymbol{k})|$ when $|x|$ is large, and are distinct near $x=0$.
	By calculating their difference, we could evaluate the GB width, which is defined as the interval where it exceeds a certain value, 
	\begin{equation}\label{width}
		\Big| \big|\hat{\phi}(x,\boldsymbol{k}) \big|-\big|\hat{\phi}_{\text{bulk}}(x,\boldsymbol{k})\big| \Big| > \alpha \max \big| \hat{\phi}_{\text{bulk}}(x,\boldsymbol{k}) \big|. 
	\end{equation}
	To unify the standard, we take $\alpha = 0.03$ for all four GBs. Their widths are $1.44a$, $1.22a$, $2.58a$, $2.74a$ for TB, NetSw, Tlt1, Tlt2, respectively. 
	Indeed, the order of GB width is consistent with that of GB energy. 
	
	\section{Conclusion}
	The mirror symmetry of TB, which is proposed in the minimal surface model \cite{chen2018minimal, han2020crystal} and roughly seen in experiments \cite{feng2021visualizing}, is affirmed in our result. 
	The dihedral angles near TB, despite may reverse the chirality, are closer to $70.5^{\circ}$ than experimental results \cite{feng2021visualizing}. 
	While TB is believed to have low GB energy due to its symmetry, we have found another GB, NetSw, having lower GB energy than TB. 
	Although NetSw does not have mirror symmetry, it has no new nodes and the two networks with reversed chirality are connected directly through struts. 
	By examining the geometric parameters describing the deformations of network, it suggests that coplanarity may be a factor more important on GB energy. 
	This is evidenced by violations of coplanarity in TB due to the formation of new nodes, and larger deviation of coplanarity in Tlt1. 
	Furthermore, inspection of spectral intensities indicates that the number of high-intensity spectra may also affect the GB energy. 
	It would be of interests to explore whether these observations are also suitable for other DG GBs, which can be addressed in future works. 
	
	\section{Author information}
	
	\textbf{Corresponding authors}
	
	\textbf{Kai Jiang} - \textit{ Hunan Key Laboratory for Computation and Simulation in Science and Engineering, Key Laboratory of Intelligent Computing and Information Processing of Ministry of Education, School of Mathematics and Computational Science, Xiangtan University, Xiangtan, Hunan, China, 411105}
	
	E-mail: kaijiang@xtu.edu.cn 
	
	\textbf{Jie Xu} - \textit{LSEC \& NCMIS, Institute of Computational Mathematics and Scientific/Engineering Computing (ICMSEC), Academy of Mathematics and Systems Science (AMSS), Chinese Academy of Sciences, Beijing, China.}
	
	E-mail: xujie@lsec.cc.ac.cn
	
	\noindent \textbf{Authors}
	
	\textbf{Jing Chen} - \textit{Hunan Key Laboratory for Computation and Simulation in Science and Engineering, Key Laboratory of Intelligent Computing and Information Processing of Ministry of Education, School of Mathematics and Computational Science, Xiangtan University, Xiangtan, Hunan, China, 411105.}
	
	\textbf{Zhangpeng Sun} - \textit{Hunan Key Laboratory for Computation and Simulation in Science and Engineering, Key Laboratory of Intelligent Computing and Information Processing of Ministry of Education, School of Mathematics and Computational Science, Xiangtan University, Xiangtan, Hunan, China, 411105}
	
	\noindent \textbf{Authors Contributions}
	
	\textbf{Jie Xu} and \textbf{Kai Jiang} designed the research, \textbf{Kai Jiang} acquired computing resources, \textbf{Jing Chen} did numerical simulations and visulization, \textbf{Jing Chen} and \textbf{Jie Xu} analyzed data, and all the authors wrote the paper.
	
	\noindent\textbf{Notes:} 
	
	The authors declare no competing financial interest.
	
	\begin{acknowledgement}
		
		This work is partially supported by the National Key Research and Development Program of China (2023YFA1008802), National Natural Science Foundation of China (12171412, 12288201, 12371414).
		We are grateful to the High Performance Computing Platform of Xiangtan University for partial support of this work.
		
	\end{acknowledgement}
	
	\begin{suppinfo}
		
		\begin{itemize}
			\item Bulk DG, Relationship of spectral indices between bulk and GB, Further information on GB structures, Figures \ref{bulk}-\ref{intensity0} and Tables \ref{tab: mode indices}-\ref{tab_tilt0}
		\end{itemize}
		
	\end{suppinfo}
	
	\bibliography{refs}
	
\end{document}

\begin{figure}[H]
	\centering
	\includegraphics[width=\textwidth]{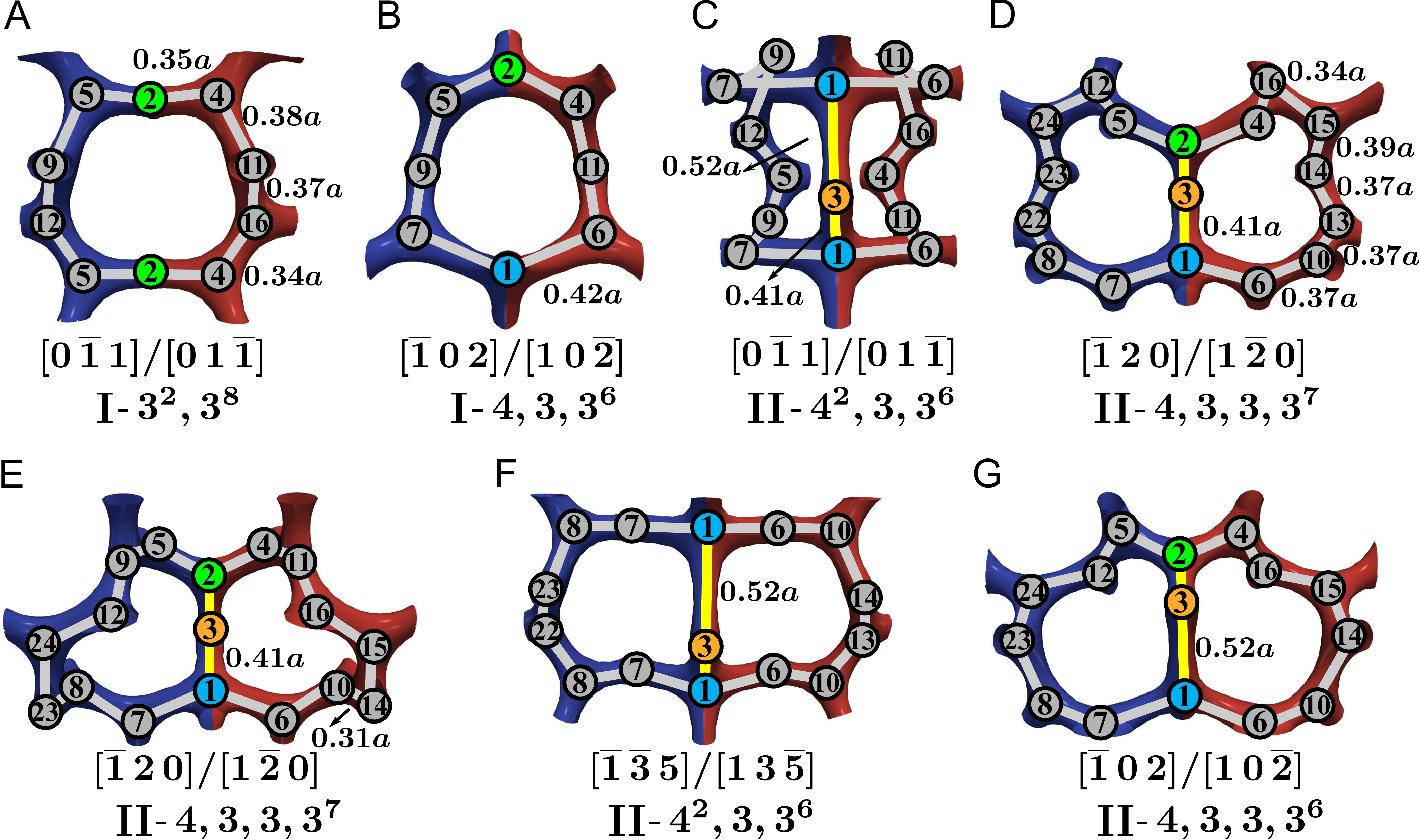}
	\caption{TB circuits and strut lengths. 
	}
	\label{twin.loop}
\end{figure}
\begin{figure}[H]
	\centering
	\includegraphics[width=\textwidth]{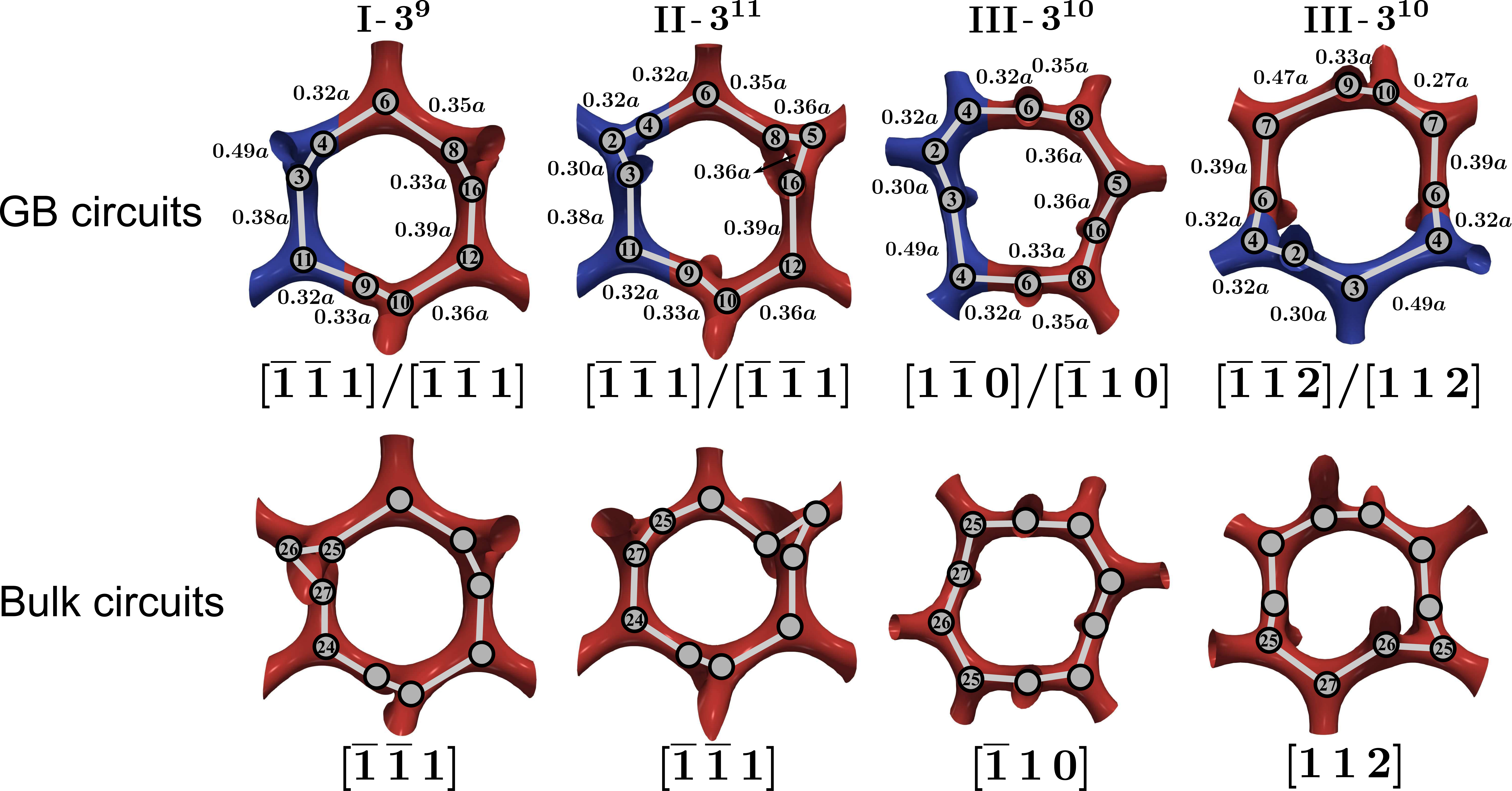}
	\caption{The circuits in NetSw and their corresponding bulk circuits in grain 2.}
	\label{hand.ring}
\end{figure}
\begin{figure}[H]
	\centering
	\includegraphics[width=\textwidth]{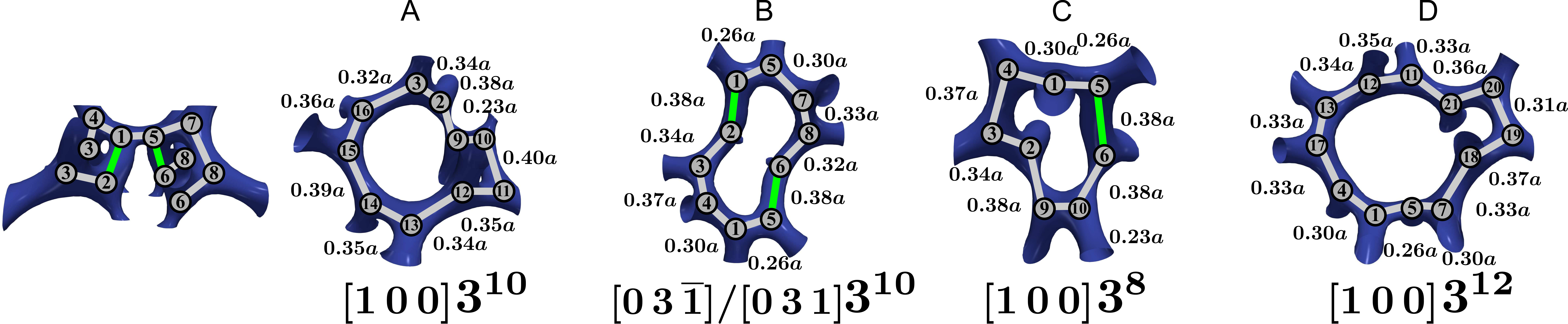}
	\caption{(A-D) Four circuits in the left-handed network of Tlt1.
	}
	\label{tilt.ring.shape.2}
\end{figure}
\begin{figure}[H]
	\centering
	\includegraphics[width=\textwidth]{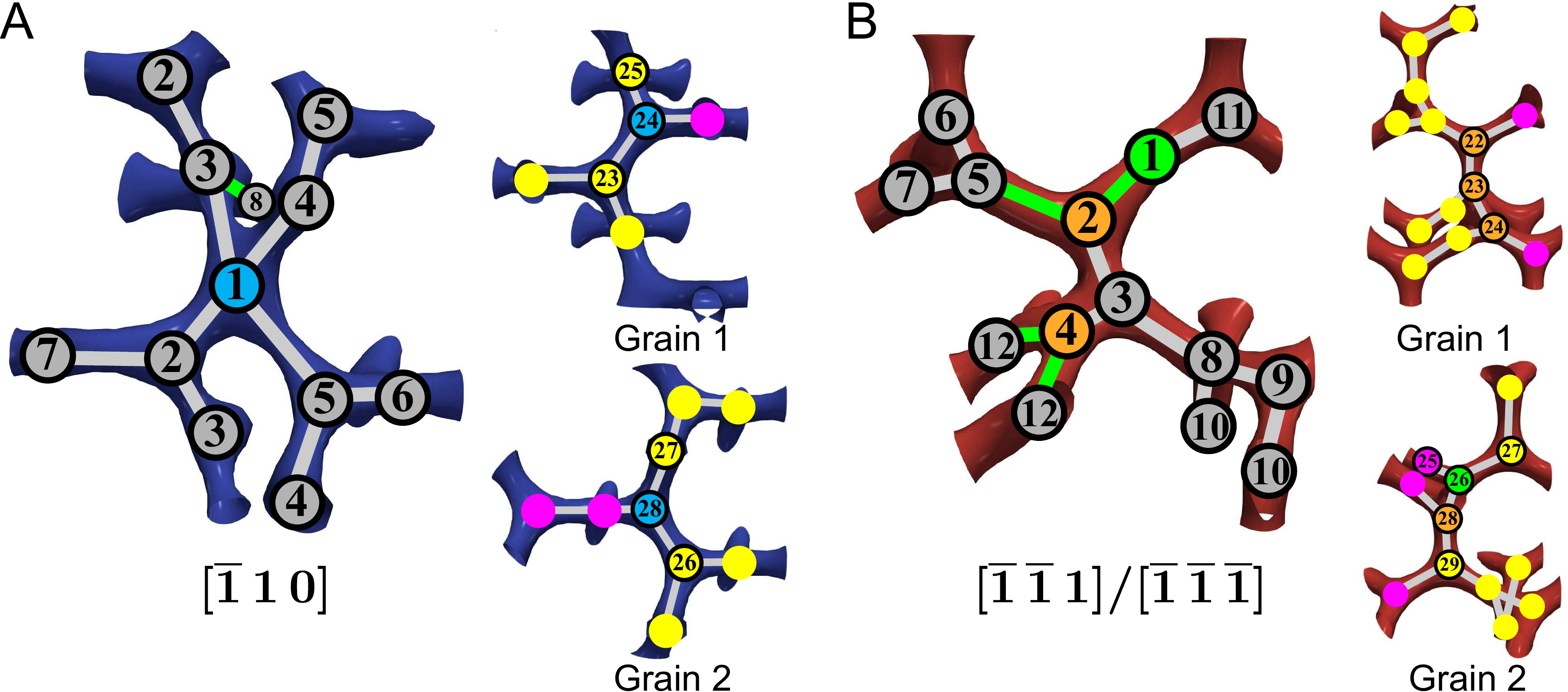}
	\caption{(A)-(B) The left- and right-handed networks neighboring new nodes in Tlt2, in comparison with bulk networks of two grains.
		The yellow nodes in two grains correspond to the gray nodes in Tlt2.
	}
	\label{tilt0.node.shape}
\end{figure}
\begin{figure}[H]
	\centering
	\includegraphics[width=\textwidth]{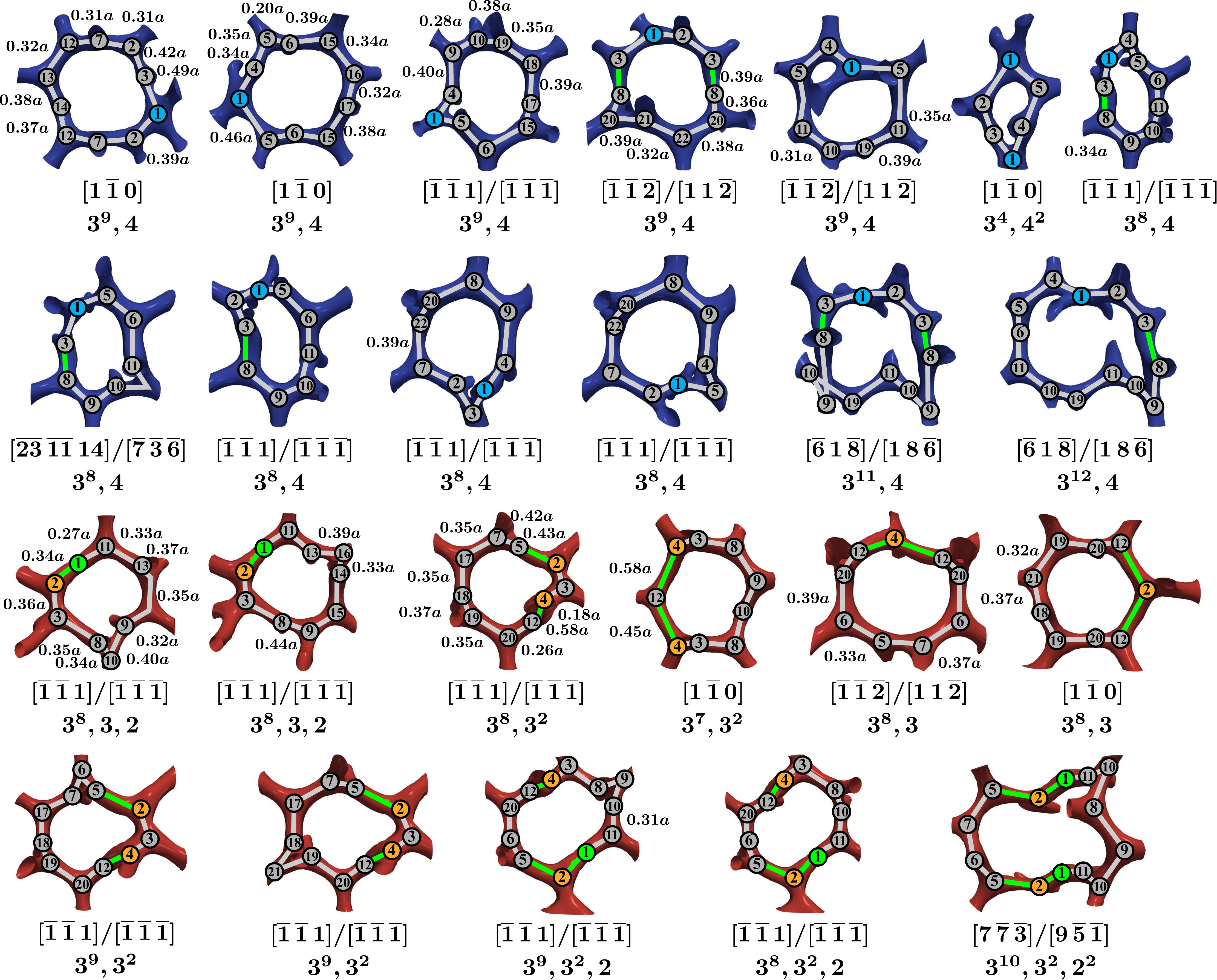}
	\caption{Circuits in Tlt2. 
	}
	\label{tilt0.ring}
\end{figure}
\begin{figure}[H]
	\centering
	\includegraphics[width=\textwidth]{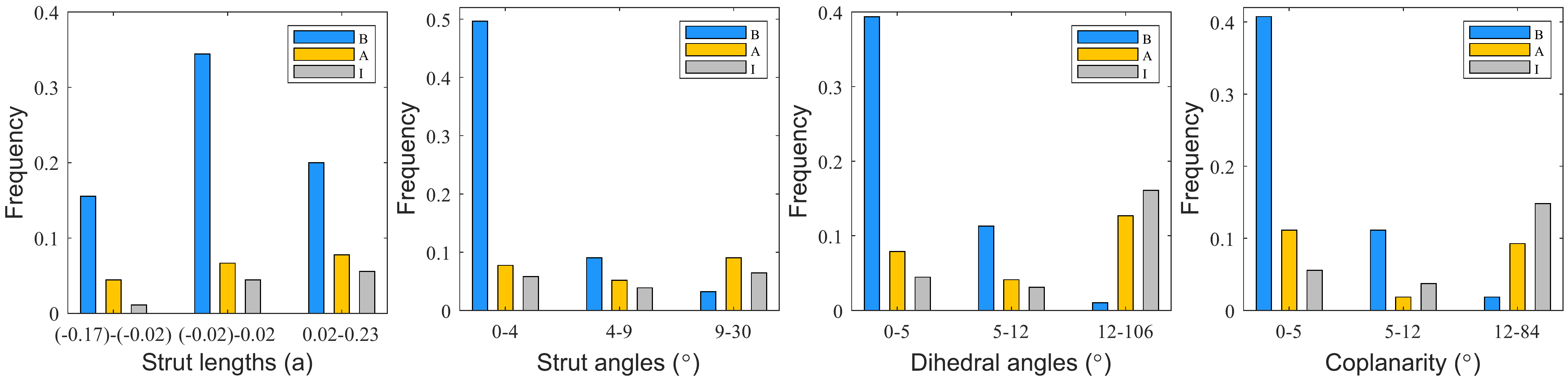}
	\caption{Statistics of deviations for strut lengths, strut angles, dihedral angles and coplanarity in Tlt2.}
	\label{data_tilt0}
\end{figure}

\setlength{\parindent}{2em}
\noindent
This PDF file includes:\\
\ref{sec:bulk} Bulk DG \\
\ref{sec:index} Relationship of spectral indices between bulk and GB\\
\ref{sec:GBs} Further information on GB structures\\
\indent Figures \ref{bulk}-\ref{intensity0} \\
\indent Tables \ref{tab: mode indices}-\ref{tab_tilt0}

\newpage

\section{Bulk DG}
\label{sec:bulk}
\begin{figure}[H]
	\centering
	\includegraphics[width=\textwidth]{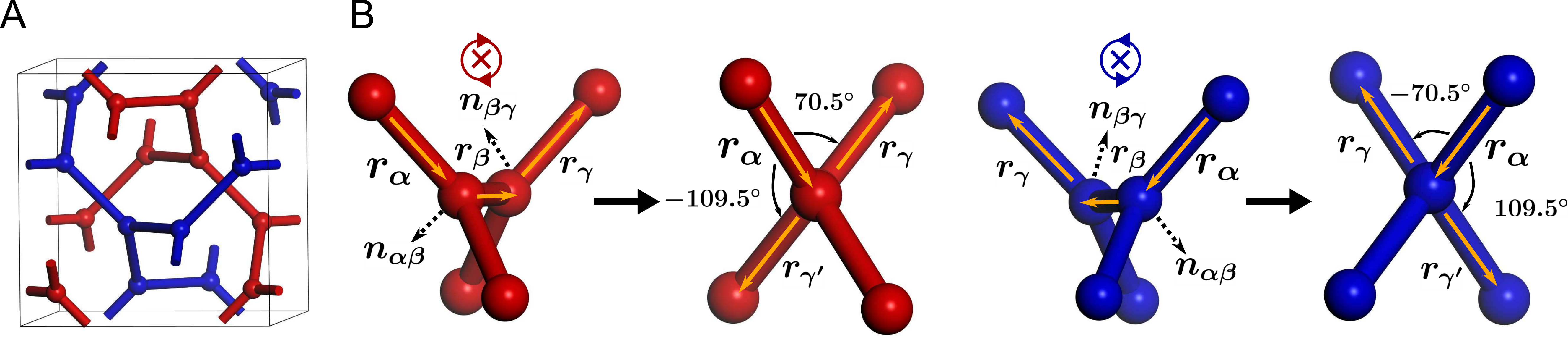}
	\caption{(A) The DG skeleton in a unit cell.
		(B) Dihedral angles in the two networks with different chiralities. 
	}
	\label{bulk}
\end{figure}

In the ideal DG skeleton (illustrated in \cref{bulk}), each node has three coplanar struts of the length $(\sqrt{2}a)/4 \approx 0.35a$ with $a$ denoting the size of a cubic unit cell. 
For three consecutive struts (\cref{bulk}\,B) whose unit vectors are $ r_\alpha$, $r_\beta$, $r_\gamma $, respectively, the dihedral angle is defined as that of two half planes containing $(r_\alpha, r_\beta)$ and $(r_\beta, r_\gamma)$, respectively, or equivalently the angle between the two normal vectors $ n_{\alpha \beta} = ( r_\alpha \times r_\beta)/\left| r_\alpha \times r_\beta \right|$ and $ n_{\beta \gamma} = (r_\beta \times r_\gamma)/\left| r_\beta \times r_\gamma \right| $.
The dihedral angles of an ideal DG are $\pm 70.5^{\circ}$ or $\mp 109.5^{\circ}$, where ``+'' (``-'') indicates clockwise (counterclockwise).

\section{Relationship of spectral indices between bulk and GB}
\label{sec:index}
The bulk DG profile can be represented by Fourier series, 
\begin{equation}
	\phi_{0}(\boldsymbol{r})=\sum_{\boldsymbol{k} \in \mathbb{Z}^{3}} \hat{\phi}_{0}(\boldsymbol{k}) e^{i(\mathcal{P} \boldsymbol{k})^{T}  \boldsymbol{r}},
\end{equation}
where $\boldsymbol{r}=(x,y,z)^T$ represents the spatial position.
For DG, the $3\times 3$ matrix $\mathcal{P}=b\mathbf{I}$ represents the primitive reciprocal lattice of the size $b=2\pi/a$. 
The integer vector $\boldsymbol{k} \in \mathbb{Z}^{3} $ is the indices of reciprocal lattice vectors.
The orientations of two grains can be expressed by two rotation matrices $R_s \in \rm{SO(3)}$, where $s =1, 2$, representing the grains in $x<0$ and $x>0$. 
Then, the profiles of two rotated grains are written as
\begin{equation}
	\begin{aligned}
		\phi_{s}(\boldsymbol{r}) &=\phi_{0}\left(R_{s} \boldsymbol{r}\right)=\sum_{\boldsymbol{k} \in \mathbb{Z}^{3}}\hat{\phi}_{0}(\boldsymbol{k}) e^{i\left(R_{s x}^{T} \mathcal{P} \boldsymbol{k}\right)x} e^{i\left(\tilde{R}_{s}^{T} \mathcal{P} \boldsymbol{k}\right)^{T}  \tilde{\boldsymbol{r}}}, \quad 
	\end{aligned}
\end{equation}
where $\tilde{\boldsymbol{r}}=(y,z)^T$, $R_s = (R_{sx},\tilde{R}_s)$. 
We define a $(2 \times 3)$-order matrix $\tilde{\mathcal{P}}_{s}=\tilde{R}_{s}^{T} \mathcal{P}$.
To construct the $y$-$z$ Fourier expansion in the GB system, we extract linearly independent column vectors from the $2 \times 6$ matrix $(\tilde{\mathcal{P}}_{1},\tilde{\mathcal{P}}_{2})$, denoted as $\tilde{\mathcal{P}}$ that is $2\times d$, such that $ \tilde{\mathcal{P}} \mathbb{Z}^{d} = (\tilde{\mathcal{P}}_{1},\tilde{\mathcal{P}}_{2})\mathbb{Z}^{6}$. In the case that the two grains have common periodicity in $y$-$z$, two vectors are sufficient i.e. $d=2$. 

The rotation matrices $R_{1}$, $R_{2}$ of two grains and the matrices $\tilde{\mathcal{P}}$ of the four GBs are provided below. 
For TB,
\begin{equation}\label{rotateTB}
	R_{1}=\left(\begin{array}{ccc}
		\frac{1}{\sqrt{6}} & -\frac{1}{\sqrt{3}} & -\frac{1}{\sqrt{2}} \\
		-\frac{2}{\sqrt{6}} & -\frac{1}{\sqrt{3}} & 0 \\
		-\frac{1}{\sqrt{6}} & \frac{1}{\sqrt{3}} & -\frac{1}{\sqrt{2}}
	\end{array}\right), \quad R_{2}=\left(\begin{array}{ccc}
		\frac{1}{\sqrt{6}} & \frac{1}{\sqrt{3}} & \frac{1}{\sqrt{2}} \\
		-\frac{2}{\sqrt{6}} & \frac{1}{\sqrt{3}} & 0 \\
		-\frac{1}{\sqrt{6}}  & -\frac{1}{\sqrt{3}} & \frac{1}{\sqrt{2}}
	\end{array}\right), \quad \tilde{\mathcal{P}}=\left(\begin{array}{cc}
		\frac{1}{\sqrt{3}}b & -\frac{1}{\sqrt{3}}b \\
		-\frac{1}{\sqrt{2}}b & 0 \end{array}\right).
\end{equation}
For NetSw, 
\begin{equation}\label{rotateNetSw}
	R_{1}=\left(\begin{array}{ccc}
		-\frac{1}{\sqrt{6}} & -\frac{1}{\sqrt{3}} & -\frac{1}{\sqrt{2}} \\
		-\frac{1}{\sqrt{6}} & -\frac{1}{\sqrt{3}} & \frac{1}{\sqrt{2}} \\
		-\frac{2}{\sqrt{6}}  & \frac{1}{\sqrt{3}} & 0
	\end{array}\right), \quad R_{2}=\left(\begin{array}{ccc}
		\frac{1}{\sqrt{6}} & -\frac{1}{\sqrt{3}} & \frac{1}{\sqrt{2}} \\
		\frac{1}{\sqrt{6}} & -\frac{1}{\sqrt{3}} & -\frac{1}{\sqrt{2}} \\
		\frac{2}{\sqrt{6}}  & \frac{1}{\sqrt{3}} & 0
	\end{array}\right), \quad \tilde{\mathcal{P}}=\left(\begin{array}{cc}
		-\frac{1}{\sqrt{3}}b & -\frac{1}{\sqrt{3}}b \\
		0 & \frac{1}{\sqrt{2}}b \end{array}\right).
\end{equation}
For Tlt1, 
\begin{equation}\label{rotateTlt1}
	R_{1}=\left(\begin{array}{ccc}
		0 & 0 & -1 \\
		-\frac{1}{\sqrt{10}} & -\frac{3}{\sqrt{10}} & 0 \\
		-\frac{3}{\sqrt{10}}  & \frac{1}{\sqrt{10}} & 0
	\end{array}\right), \quad R_{2}=\left(\begin{array}{ccc}
		0 & 0 & -1 \\
		\frac{1}{\sqrt{10}} & -\frac{3}{\sqrt{10}} & 0 \\
		-\frac{3}{\sqrt{10}}  & -\frac{1}{\sqrt{10}} & 0
	\end{array}\right), \quad \tilde{\mathcal{P}}=\left(\begin{array}{cc}
		\frac{1}{\sqrt{10}}b & 0 \\
		0 & \frac{3}{\sqrt{10}}b \end{array}\right).
\end{equation}	
For Tlt2, 
\begin{equation}\label{rotateTlt2}
	R_{1}=\left(\begin{array}{ccc}
		-\frac{1}{\sqrt{6}} & -\frac{1}{\sqrt{3}} & -\frac{1}{\sqrt{2}} \\
		-\frac{1}{\sqrt{6}} & -\frac{1}{\sqrt{3}} & \frac{1}{\sqrt{2}} \\
		-\frac{2}{\sqrt{6}} & \frac{1}{\sqrt{3}} & 0
	\end{array}\right), \quad R_{2}=\left(\begin{array}{ccc}
		\frac{1}{\sqrt{6}} & -\frac{1}{\sqrt{3}} & -\frac{1}{\sqrt{2}} \\
		\frac{1}{\sqrt{6}} & -\frac{1}{\sqrt{3}} & \frac{1}{\sqrt{2}} \\
		-\frac{2}{\sqrt{6}} & -\frac{1}{\sqrt{3}} & 0
	\end{array}\right), \quad \tilde{\mathcal{P}}=\left(\begin{array}{cc}
		\frac{1}{\sqrt{3}}b & -\frac{1}{\sqrt{3}}b \\
		0 & \frac{1}{\sqrt{2}}b \end{array}\right).
\end{equation}	

We take TB as an example to demonstrate how to find the relationship between GB's and DG's spectral indices. 
For TB, we have
\begin{equation}
	\tilde{\mathcal{P}}_{1}=\tilde{R}_{1}^{T} \mathcal{P}=\left(\begin{array}{ccc}
		-\frac{1}{\sqrt{3}}b & -\frac{1}{\sqrt{3}}b & \frac{1}{\sqrt{3}}b \\
		-\frac{1}{\sqrt{2}}b & 0 & -\frac{1}{\sqrt{2}}b 
	\end{array}\right), \quad \tilde{\mathcal{P}}_{2}=\tilde{R}_{2}^{T} \mathcal{P}=\left(\begin{array}{ccc}
		\frac{1}{\sqrt{3}}b & \frac{1}{\sqrt{3}}b & -\frac{1}{\sqrt{3}}b\\
		\frac{1}{\sqrt{2}}b  & 0 & \frac{1}{\sqrt{2}}b
	\end{array}\right),
\end{equation}
For 2D index $\boldsymbol{k} = (k_a,k_b)^{T}\in\mathbb{Z}^2$, the spectrum vector $(k_y, k_z)^{T}$ of TB is
\begin{equation}\label{eq:twin_real_spectra}
	\tilde{\mathcal{P}} \boldsymbol{k} =\left(\begin{array}{c}
		\frac{1}{\sqrt{3}}b(k_a-k_b) \\
		-\frac{1}{\sqrt{2}}bk_a  \end{array}\right).
\end{equation}
For 3D index $\boldsymbol{k}' = (k_1,k_2,k_3)^{T}\in\mathbb{Z}^3$, 
the spectrum vector $(k_y, k_z)^{T}$ of grain 2	is
\begin{equation}\label{eq:right_bulk_real_spectra}
	\tilde{\mathcal{P}}_2 \boldsymbol{k}' = \left(\begin{array}{c}
		\frac{1}{\sqrt{3}}b(k_1+k_2-k_3) \\
		\frac{1}{\sqrt{2}}b(k_1+k_3)  \end{array}\right).
\end{equation}
Then we can obtain the relationship between TB and bulk grain 2. For other GBs, the procedure is similar. \cref{tab: mode indices} summarizes the relationship.

\section{Further information on GB structures}\label{sec:GBs}
\subsection{TB}
\label{appendix:TB}
\begin{figure}[H]
	\centering
	\includegraphics[width=\textwidth]{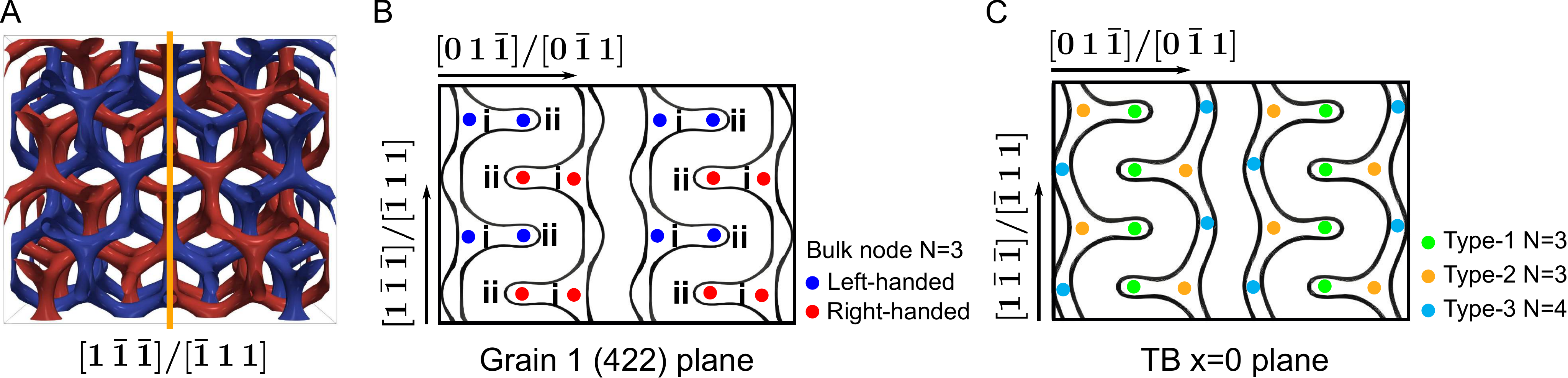}
	\caption{(A) The TB skeleton, viewed along $[1\bar{1}\bar{1}]$ (grain 1) and $[\bar{1}11]$ (grain 2) directions.
		(B) The $(422)$ slice in grain 1 and the nodes. 
		(C) The $x=0$ plane in TB and the nodes. 
	}
	\label{twin.real}
\end{figure}

The TB skeleton is viewed from a different direction in \cref{twin.real}\,A. 
Comparing the $(422)$ slice of grain 1 (\cref{twin.real}\,B) and $x=0$ slice of TB (\cref{twin.real}\,C), 
we could see that Type-1 to Type-3 nodes are formed exactly in the way claimed in Ref.\,\cite{feng2021visualizing}. 
The circuits containing TB nodes not given in the main text are drawn in \cref{twin.real}. 

\subsection{NetSw}
\label{appendix:rotation-CCB}
The four circuits passing $x=0$ with more red nodes are presented in \cref{hand.ring}. 
Still, we compare these circuits with bulk circuits of grain 2. It is observed that the positions of blue nodes 11 and 4 are close to red nodes 24 and 25 in the bulk, respectively, with connections distinct from the bulk. 
This generates circuits of different lengths. 

\subsection{Tlt1}
\label{appendix:rotation-tilt}
\cref{tilt.ring.shape.2} illustrates two curved struts and four circuits in the blue network of Tlt1.
Circuits crossing $x=0$ are formed by combining part of grain 1 and grain 2 at edges 1-5, 9-10, and 11-12.
Strut lengths involving these nodes (node 1, 5, 9, 10) change larger than others relative to bulk strut length.

\subsection{Tlt2}
\label{appendix:rotation-tilt0}
\begin{figure}[H]
	\centering
	\includegraphics[width=\textwidth]{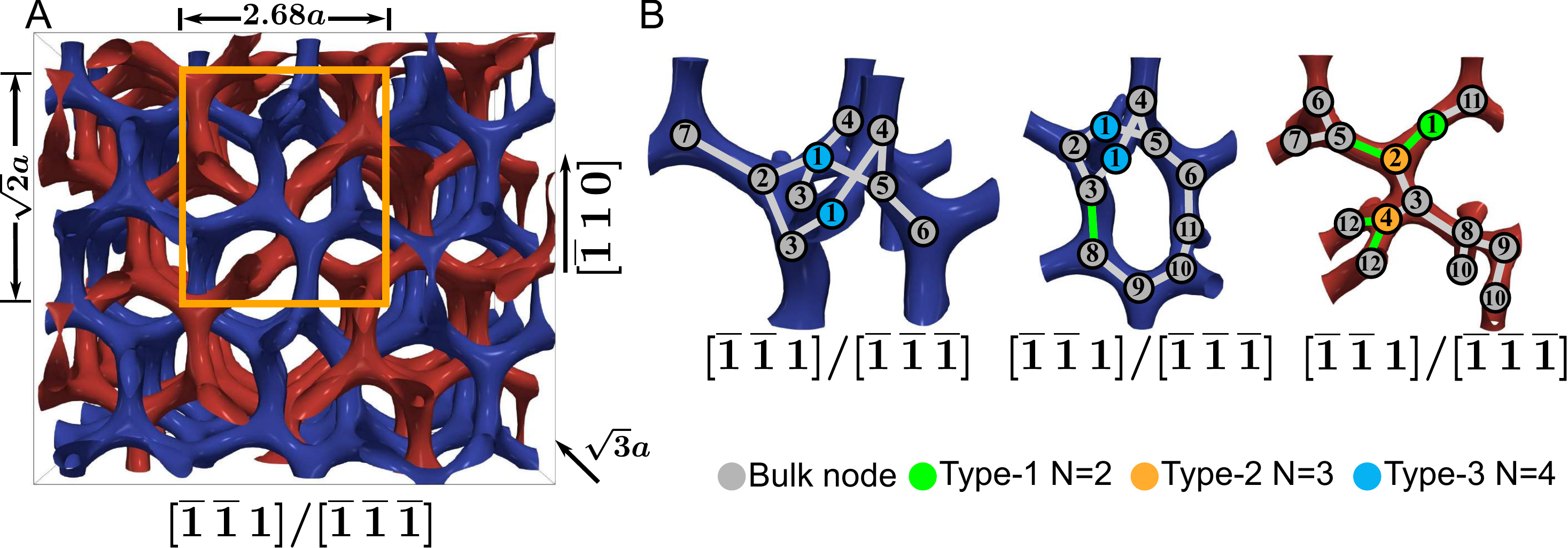}
	\caption{(A) Tlt2 skeleton. 
		The orange box spans one period in the $y$-direction.
		(B) Three new nodes marked with different colored dots and five curved struts marked with green lines.
	}
	\label{tilt0.real}
\end{figure}

\cref{tilt0.real}\,A illustrates the Tlt2 skeleton, formed by the connection of two grains with identical chirality network at $x=0$.
The period in the $y$-$z$ plane is $\sqrt{3}a \times \sqrt{2}a$.
Three new nodes and five curved edges emerge within an orange box spanning approximately $2.68a$ in the $x$-direction, as shown in \cref{tilt0.real}\,B.
The nodes in the blue and red networks are numbered independently.
To understand these new nodes, we compare them to the corresponding bulk networks of two grains (\cref{tilt0.node.shape}).
In the left-handed network, node 1 with four edges (Type-3) results from the fusion of nodes 24 and 28, each with two edges. 
Similarly, in the right-handed network, node 2 also emerges as a merging of two bulk DG nodes from two grains (node 22 and 28).
Node 4 with three edges (Type-2) is formed by the fusion of two bulk DG nodes of grain 1 (node 23 and 24).
Moreover, the strut 25-26 of grain 2 is truncated, forming a new node with two edges (Type-1).
Tlt2 circuits crossing $x=0$ (\cref{tilt0.ring}) are formed in the blue network by combining parts of grain 1 and grain 2 at edges 1-4, 1-5, 8-9, and in the red network at edge 2-3.
In the blue network, strut lengths involving nodes 1, 4, 5, 8, 9 connecting two grains, as well as in the red network involving nodes 1, 2, 3, deviate larger than others.

\subsubsection{Deformation statistics}\label{sec:tilt0_shape}
The deformation statistics of four geometric parameters in Tlt2 within the range of $x\in[-1.58a, 1.58a]$ are presented in \cref{data_tilt0} and \cref{tab_tilt0}.
The four geometric parameters of Tlt2 are examined in region I, A and B. 
Tlt2 deforms more than the other three GBs in all four aspects, potentially leading to a higher GB energy.
The substantial deviations in Tlt2 occur mainly in Region I, especially in dihedral angles.
In region A and B, strut lengths and dihedral angles (exceeding $66\%$) are more prone to deformation than coplanarity (within $16\%$) and strut angles (within $21\%$).
Specifically, dihedral angles seem to deviate the most, while coplanarity is comparatively well-maintained, aligning with our expectations.

\subsubsection{Spectral configurations}
\begin{figure}[H]
	\centering
	\includegraphics[width=\textwidth]{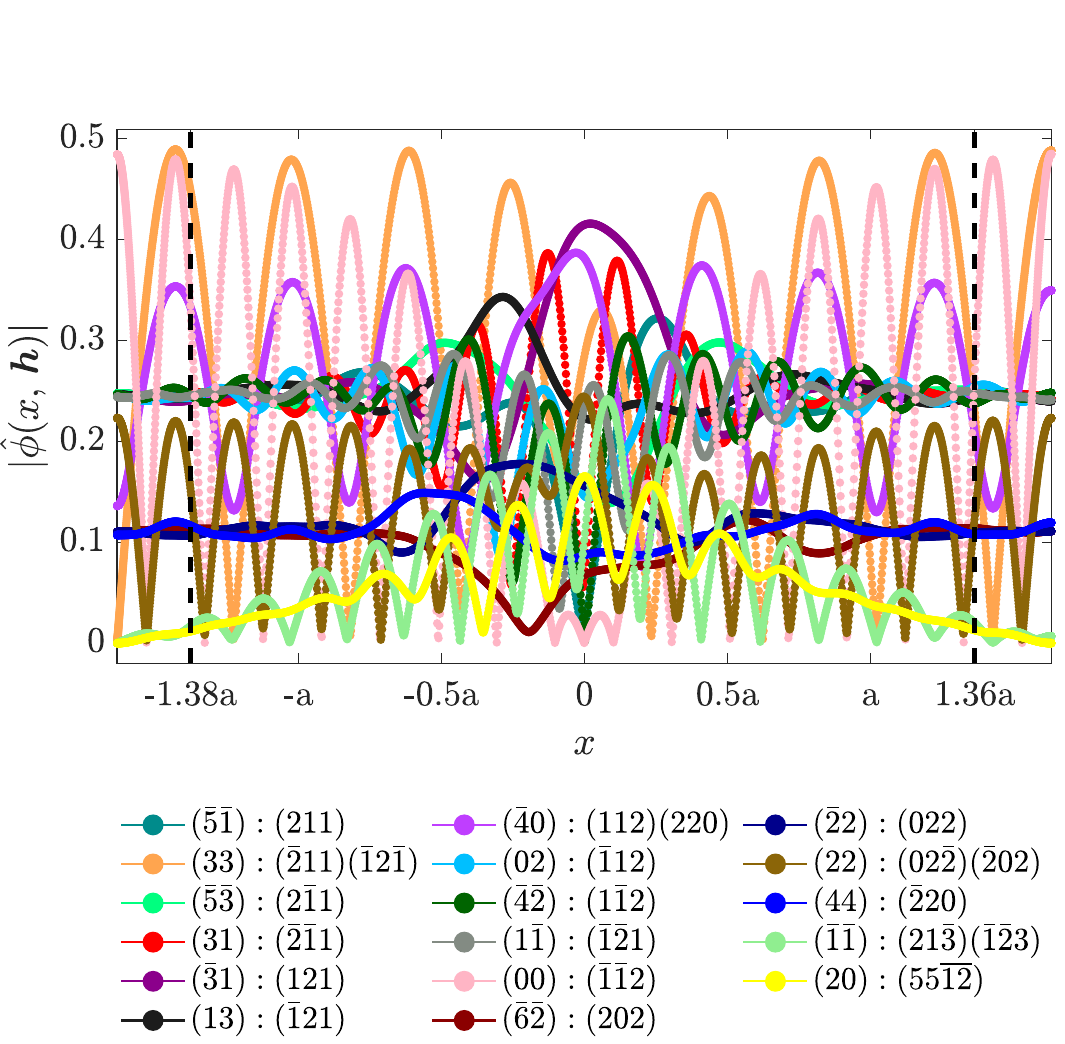}
	\centering
	\caption{Intensities of spectra against $x$ in Tlt2, for those maximum over $0.09$.
		The black dotted lines represent the evaluated boundary of GB with $\alpha = 0.03$.
	}
	\label{intensity0}
\end{figure}

In Tlt2, there are $17$ spectra with the intensities exceeding $0.09$ (\cref{intensity0}), while others have the intensities sufficiently smaller. 
Of these spectra, not only are the projections of major bulk spectra present, but also two others with maxima exceeding $0.1$, potentially contributing to higher GB energy.
Based on \cref{width}, the GB width of Tlt2 is $2.74a$, exceeding that of the other three GBs.

\begin{table}[H]
	\centering
	\caption{The relationship between GB's and DG's spectral indices.}
	\begin{tabular}{ccc}
		\toprule
		& $k_y$ & $k_z$ \\
		\midrule
		TB & $k_a - k_b = k_1+k_2-k_3$ & $k_a = -k_1-k_3$ \\
		NetSw & $k_a + k_b = k_1+k_2-k_3$ & $k_b = k_1-k_2$ \\
		Tlt1 & $k_a = -3k_2 - k_3$ & $k_b = -k_1$\\
		Tlt2 & $k_a - k_b = -k_1 - k_2 - k_3$ & $k_b = -k_1 + k_2$ \\
		\bottomrule
	\end{tabular}
	\label{tab: mode indices}
\end{table}

\begin{table}[H]
	\centering
	\caption{Maximum variations for strut lengths, dihedral angles, strut angles and coplanarity in Tlt2.}
	\begin{tabular}{cccccc}
		\toprule
		\multirow{2}{*}{structure} & \multirow{2}{*}{region} & strut length & strut angle  &  dihedral angle & coplanarity \\
		&  & $(0.35a)$ & $(120^{\circ})$  &  $(\pm70.5^{\circ})$ & $(180^{\circ})$\\
		\midrule
		\multirow{3}{*}{Tlt2} & I & $\pm0.17a$ & $ \pm 30^{\circ}$ &$ \pm 102^{\circ}$& $\pm 84^{\circ}$ \\
		& A & $\pm0.23a$ & $ \pm 25^{\circ}$ &$ \pm 106^{\circ}$& $\pm 29^{\circ}$ \\
		& B & $\pm0.09a$ & $ \pm 16^{\circ}$ & $ \pm 12^{\circ}$  & $\pm 14^{\circ}$ \\
		\bottomrule
	\end{tabular}
	\label{tab_tilt0}
\end{table}

\bibliography{refs}